# THREE TIER ENCYRPTION ALGORITHM FOR SECURED FILE TRANSFER

Bhargav.Balakrishnan
Systems Engineer, Bahwan CyberTek LLC
Muscat, Oman
Sherkhan_sastra@yahoo.co.in

*Abstract:* **This encryption algorithm is mainly designed for having a secure file transfer in the low privilege servers and as well as in a secured environment too. This methodology will be implemented in the data center and other important data transaction sectors of the organisation where the encoding process of the software will be done by the database administrator or system administrators and his trusted clients will have decoding process of the software. This software will not be circulated to the unauthorised customers.**

## I. INTRODUCTION

This algorithm comprises of RSA (2048 Bits), number conversions (Binary, Octal etc...), digital encoding (4B/5B, 8B/10B block encoding), mathematical series (Sine, Cosine, Exponential, Tan etc...). All these process will provide a secured encrypted output of the data which will prevent the hacker from viewing the data when they are transmitted over the network at the same even if the hacker gets the encrypted data it will be many thousands of years to get a clue upon it. This type of algorithm is needed for any organisation which is need of having a secured data transmission within their network.

## II. ALGORITHM

The steps of using this encryption methodology are as follows

### A. RSA Algorithm

In the first step the RSA algorithm will be carried with the following modifications

a) Consider two prime numbers as 11 and 13 – (1)
b) N= P *Q i.e. 143
c) M = (P-1)(Q-1) i.e. 120
d) D is the decryption key Example 3 which is a prime number
e) E = D inverse (mod n) i.e. 47
f) Let the password be "Hello "take the ASCII value of the password covert it as 7269767679.
g) Concatenate this ASCII value with a SALT value (Randomly generated number) say 34 i.e. 247172101086
h) Finally multiply this with the Encryption value to get final encrypted word 9886884043440

Modification and Requirement in RSA Algorithm

- The minimum requirement for P and Q values in RSA is 2048 bits which gives the utmost security to the file that is being transferred.
- Modification is inclusion of ASCII value conversion and SALT Value. Here SALT is being left **user defined**.
- The P and Q values are also **user defined** that is also a modification
- At this you can use any encryption algorithms which are being updated.

### B. Number conversion

The above arrived result through RSA – **9886884043440** will be converted into 0's and 1's using number conversion.

The above **encrypted data** (9886884043440) will be converted as into any of the following conversions like **hexadecimal/octal/binary** i.e. 1001011011011100101001001010100110110000100101 1011011001010010010010110110111100101001001010 1001101100001001011011011100101001001010100110 110000. This is for binary in the same way it can be done for octal /hexadecimal

### C. Digital Encoding

This number conversion will be modified using Digital Encoding (Either Line or Block Encoding). Advantage: - Rearranges the bits of data i.e. 0's and 1's.

| 4 bit value nibble | 5 bit value symbol | 4 bit value nibble | 5 bit value symbol |
|---|---|---|---|
| 0000 | 11110 | 1000 | 10010 |
| 0001 | 01001 | 1001 | 10011 |
| 0010 | 10100 | 1010 | 10110 |
| 0011 | 10101 | 1011 | 10111 |
| 0100 | 01010 | 1100 | 11010 |
| 0101 | 01011 | 1101 | 11011 |
| 0110 | 01110 | 1110 | 11100 |
| 0111 | 01111 | 1111 | 11101 |

Fig. 1 Diagram of 4B/5B Substitution Block Encoding

Then use any of the **line encoding schemes** like **NRZ, NRZ-I ,RZ, biphase (Manchester, and differential Manchester), AMI and pseudo ternary, 2B /IQ , 8B/6T, and 4d –PAMS and MLTS** that will convert the number which are being as binary in the above as follows

Let us consider **4B/5B Block Encoding**

10011011101101111010101100101010110100111011111110100110111011011110101011001010100110111011011110101011001010101101001110111111010011011101101111010101100101010110100111011111110

*D.  Reverse process of Number Conversion*
In this step the conversions of data into 4B/5B will be converted back into numbers using number conversions. This is reverse process of Step 2Conversion back to **binary** will be give different encrypted word because of the usage of **4B/5B** line encoding. The solution will be **10205099**. Here also the **conversion** can be any one of the following **binary/octal/hexadecimal**.

*E.  Mathematical Series*
In this step the above obtained number in step 4 10205099 will be considered as the X Value. This will be substituted in the Mathematical Series.

Here in the example the **sine series** is being used.

Formulae: -   $\sin(x) = X - X^3/3! + X^5/5! - \ldots$

Here X is the encrypted value 10205099. The series is used defined say N=3 then the series will be till $X^7/7!$. Then the final result will be 10205099 - 177133382601080597549.833333 + 246018586945945274.37476851851852 -

176887364014124447176.45856481478. Use the round off function to get the final encrypted word as **176887364014124447176.**

### III.  ADVANTAGE OF USING DIGITAL ENCODING, NUMBER CONVERSION AND MATHEMATICAL SERIES

- The main advantage of Step 2, 3, 4 is in Step 2 the encrypted data obtained by RSA is converted into 0's and 1's. Then by using Digital Encoding the rearrangement of Bit's are done. Finally in Step 4 the reverse process of number conversion. **What it does?** The hacker will never get a clue of this process that is being carried unless he gets an idea about this algorithm.
- Then Step 5 also a vital role as here the number X i.e. the value obtained from Step 4 has to be determined by the hacker, for which he should what is used, if found what mathematical series used which will takes ages to refine.

But for an organisation to encrypt and decrypt will be a simple as the process involved in each data encryption will be stored in their database. So this twist in the algorithm will be playing the most important in preventing the hacking of data's.

How this methodology gives utmost security to the file at the same time increases the complexity in identifying the content by the intruder. These are being described below

If the Intruder gets this encrypted word the following things are to be determined. Determining those values is a long process and finding those will take many years in order to arrive at the conclusion

- The value of N i.e. the length of the series has to be determined.
- After finding N values the value of X has to be determined that has been substituted in the series
- In the line encoding process the split up of the bits has to be determined like 4 bits, 8 bits and so on.
- After determining this, the type of encoding has to be determined and the substitution used as in the B8SZ where 8 bit value is substituted in place of continuous 8 zero's.
- Based upon which the entire two stages can be revealed from this the first stage can be proceeded that is RSA instead of that AES, SHA, MD5 any encryption algorithm can be used.The speciality of

RSA is in determining the prime numbers P and Q which itself will take many years to determine.

The end user can be a data center, search engine etc which will get utmost security because of the usage of Line Encoding and Mathematical series. The line coding will convert the original encrypted word into duplicate encrypted word by using the following) i) binary/octal/hexadecimal the encrypted word is converted as 0's and 1's ii) then line encodings is used. This will act as a protection. This will be even more protective by using the mathematical series. On the whole the methodology will be a secure path for the transfer of data's.Time for generating the Encrypted file using this method will be comparatively less in the high end PC's with dual core processor and above with 2GB RAM with processor speed of 2.2 GHz. The RSA encryption of about 2048 bits will take time other steps will take fraction of seconds for generating the desired output.

### IV. OBSERVATION – TIME REQUIREMENT TO GENERATE ENCRYPTED DATA

- File Size : 11 KB
- No of words (Including symbols): 50
- Total time take for RSA (2048 bits): 45secs
- For other process hardly: 30secs

### V. RECOMMENDED SERVER CONFIGURATION

HP Proliant DL320 G5

- Minimum HDD Capacity 72 GB ( 2 SAS HDD)
- RAID 0+1 ( Mirroring) as a redundancy
- 2GB RAM
- Dual Core Xeon Processor
- Processor Speed 1.73 GHz and above

Why this specification?

Because while encrypting multiple data's at the same time this hardware configuration will process the entire encrypting operation in less duration.

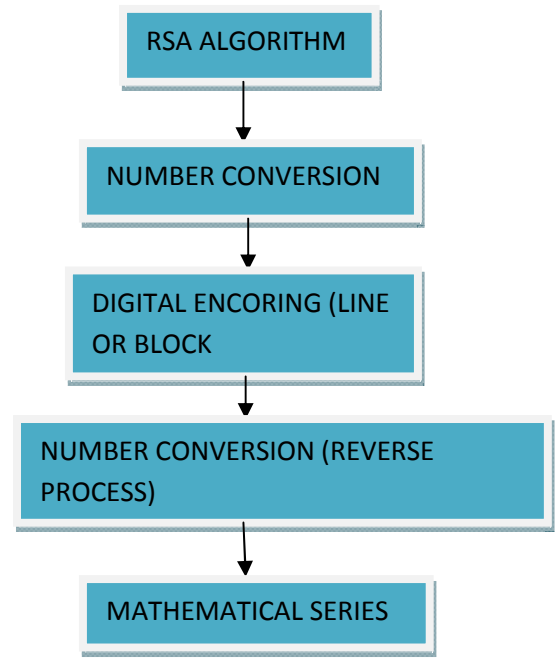

Fig.2 Diagrammatic Representation of Entire Data Encryption Process

### VI. ADVANTAGES OF THIS ENCRYPTION METHODOLOGY

- In the file transfer preferably in the low privilege servers which are an endangered place of hackers.
- In the WAN where the data transfer is not that secured, in order to give a firm security this methodology can be adopted.
- This methodology will be of high value in the defence sector where security is given high preference. Using this methodology the hacker will not be able to trace the ideas unless or until he is well versed in the mathematical and electrical technique of disclosing the data.
- This will also play a vital role in other sectors like Bank, IT, Aero Space and many more where the data transfer is given more security.

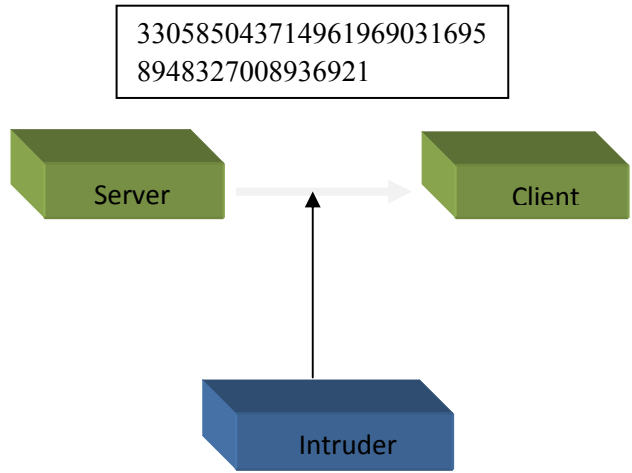

Fig.3 Diagrammatic Representation of an Intruder hacking data

As per the above diagram, we all know the process involved in the hacking of data. The encrypted data will be in the form 0's and 1's during transmission. So there will be a special feature in this algorithm.

Yes there is a special feature behind this algorithm. This algorithm will be framed as software with the same pattern flow and will be given to authorised users or organisation. What are criteria's involved in the Software

- The software will be designed using Java and MySQL as both are feasible technically, operationally etc...
- The software will be their only with administrators (System admin, data admin etc...) of the organisation.
- The software will be loaded on the server and will be configured for encrypting and decrypting process
- Flow of the software process is shown

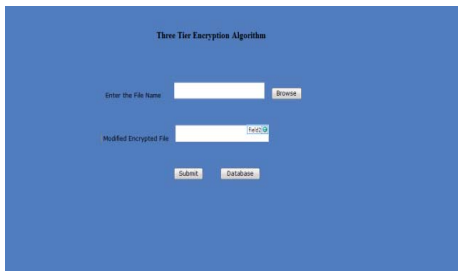

Fig.4 Homepage

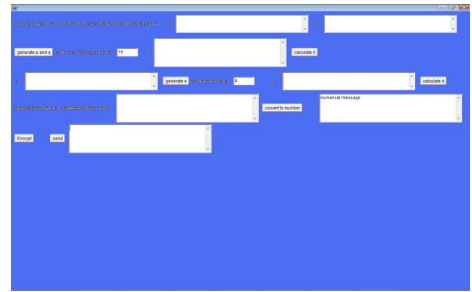

Fig .5 RSA Algorithm

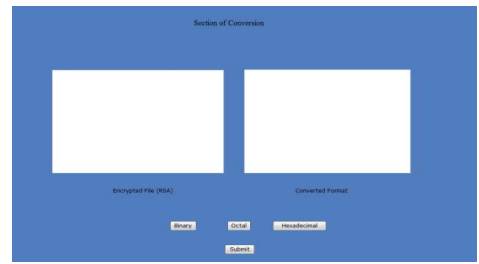

Fig.6 Number Conversion

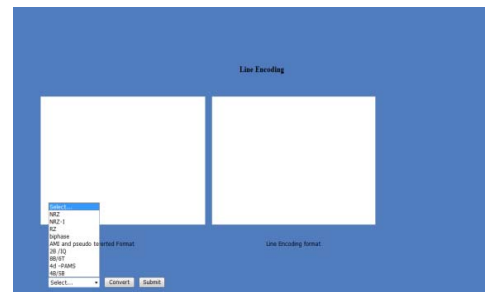

Fig.7 Digital Encoding

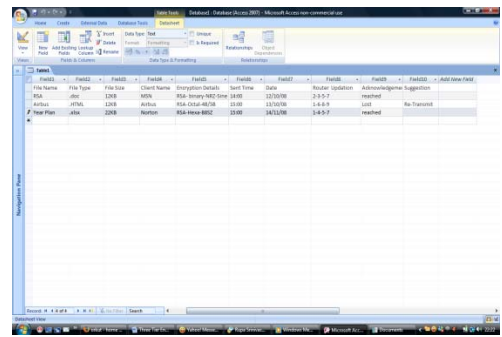

Fig.8 Database Storage for Encrypted Data

The same way the decoding process will also be carried out. The screenshots are designed using Microsoft

FrontPage to show how this application should be customized.

## VII. PRESENT APPROACH

In this present world, many organisations go with the third party like VeriSign etc. They adopt certain policies to give security for the data transaction over the network. But along with that if initially the system administrator do this encryption process with this application the organisation can be even more assured of their data being secured.

## VIII. CONCLUSION

This methodology will give the security to the files of any sizes and can be transmitted over long distances i.e. WAN. The server will have the encryption methodology and decryption process will be specified to the trusted parties/clients. Finally I want to mention the main complexity it creates for the hacker is in the understanding the type of operations used here. This methodology will definitely be an effective way to secure the files over the low privilege, which is the main goal of this project.